\documentclass[letters, usenatbib]{mnras}

\usepackage{graphicx}
\usepackage{natbib}
\usepackage{color}

\def\hi{\relax \ifmmode {\mbox H\,{\scshape i}}\else H\,{\scshape i}\fi}

\def\hii{\relax \ifmmode {\mbox H\,{\scshape ii}}\else H\,{\scshape ii}\fi}

\def\nii{\relax \ifmmode {\mbox N\,{\scshape ii}}\else N\,{\scshape ii}\fi}

\def\oi{\relax \ifmmode {\mbox O\,{\scshape i}}\else O\,{\scshape i}\fi}

\def\oii{\relax \ifmmode {\mbox O\,{\scshape ii}}\else O\,{\scshape ii}\fi}

\def\oiii{\relax \ifmmode {\mbox O\,{\scshape iii}}\else O\,{\scshape iii}\fi}

\def\cii{\relax \ifmmode {\mbox C\,{\scshape ii}}\else C\,{\scshape ii}\fi}

\def\ciii{\relax \ifmmode {\mbox C\,{\scshape iii}}\else C\,{\scshape iii}\fi}

\def\civ{\relax \ifmmode {\mbox C\,{\scshape iv}}\else C\,{\scshape iv}\fi}

\def\hei{\relax \ifmmode {\mbox He\,{\scshape i}}\else He\,{\scshape i}\fi}

\def\heii{\relax \ifmmode {\mbox He\,{\scshape ii}}\else He\,{\scshape ii}\fi}

\def\mgii{\relax \ifmmode {\mbox Mg\,{\scshape ii}}\else Mg\,{\scshape ii}\fi}

\def\sii{\relax \ifmmode {\mbox S\,{\scshape ii}}\else S\,{\scshape ii}\fi}

\def\neiii{\relax \ifmmode {\mbox Ne\,{\scshape iii}}\else Ne\,{\scshape iii}\fi}

\def\ariv{\relax \ifmmode {\mbox Ar\,{\scshape iv}}\else Ar\,{\scshape iv}\fi}

\def\ni{\relax \ifmmode {\mbox N\,{\scshape i}}\else N\,{\scshape i}\fi}

\def\ariii{\relax \ifmmode {\mbox Ar\,{\scshape iii}}\else Ar\,{\scshape iii}\fi}

\def\caii{\relax \ifmmode {\mbox Ca\,{\scshape ii}}\else Ca\,{\scshape ii}\fi}

\title[Stellar age \& outer light profiles in spirals]{No direct coupling between bending of galaxy disc stellar age and light profiles} %\thanks{Based on observations collected at the Centro Astron\'omico Hispano Alem\'an (CAHA) at Calar Alto, operated jointly by the Max-Planck Institut f\"ur Astronomie and the Instituto de Astrof\'isica de Andaluc\'ia (CSIC).}}

\author[T. Ruiz-Lara et al.]{T. Ruiz-Lara,$^{1, 2}$\thanks{E-mail: ruizlara@ugr.es (TRL)}  I. P\'erez,$^{1, 2}$  E. Florido,$^{1, 2}$  P. S\'anchez-Bl\'azquez,$^{3}$  J. M\'endez-Abreu,$^{4}$  \newauthor  M. Lyubenova,$^{5}$ J. Falc\'on-Barroso,$^{6, 7}$ L. S\'anchez-Menguiano,$^{1, 8}$ S. F. S\'anchez,$^{9}$  \newauthor  L. Galbany,$^{10, 11}$ R. Garc\'ia-Benito,$^{8}$ R. M. Gonz\'alez Delgado,$^{8}$ B. Husemann,$^{12}$ \newauthor C. Kehrig,$^{8}$  \'Angel R. L\'opez-S\'anchez $^{13, 14}$ R. A. Marino,$^{15, 16}$ D. Mast,$^{17, 18}$ P. \newauthor Papaderos,$^{19}$ G. van de Ven,$^{20}$ C.J. Walcher,$^{21}$ S. Zibetti,$^{22}$ and the CALIFA team \\
$^{1}$ Departamento de F\'isica Te\'orica y del Cosmos, Universidad de Granada, Campus de Fuentenueva, E-18071 Granada, Spain \\
$^{2}$ Instituto Carlos I de F\'isica Te\'orica y computacional, Universidad de Granada, E-18071 Granada, Spain \\ 
$^{3}$ Departamento de F\'isica Te\'orica, Universidad Aut\'onoma de Madrid, E-28049 Cantoblanco, Spain \\ 
$^{4}$ School of Physics and Astronomy, University of St Andrews, SUPA, North Haugh, KY16 9SS St Andrews, UK \\ 
$^{5}$ Kapteyn Astronomical Institute, University of Groningen, PO Box 800, NL-9700 AV Groningen, the Netherlands \\
$^{6}$ Instituto de Astrof\'isica de Canarias, Calle V\'ia L\'actea s/n, E-38205 La Laguna, Tenerife, Spain \\
$^{7}$ Departamento de Astrof\'isica, Universidad de La Laguna, E-38200 La Laguna, Tenerife, Spain \\
$^{8}$ Instituto de Astrof\'isica de Andaluc\'ia (CSIC), Glorieta de la Astronom\'ia s/n, Aptdo. 3004, E-18080 Granada, Spain \\
$^{9}$ Instituto de Astronom\'ia, Universidad Nacional Aut\'onoma de M\'exico, A.P. 70-264, 04510 M\'exico D.F., Mexico\\
$^{10}$ Millennium Institute of Astrophysics, Santiago, Chile \\
$^{11}$ Departamento de Astronom\'ia, Universidad de Chile, Camino El Observatorio 1515, Las Condes, Santiago, Chile \\
$^{12}$ European Southern Observatory (ESO), Karl-Schwarzschild-Str. 2, D-85748 Garching b. M\"unchen, Germany \\
$^{13}$\,Australian Astronomical Observatory, PO Box 915, North Ryde, NSW 1670,Australia\\
$^{14}$\,Department of Physics and Astronomy, Macquarie University, NSW 2109, Australia\\
$^{15}$ CEI Campus Moncloa, UCM-UPM, Departamento de Astrof\'sica y CC. de la Atm\'osfera, Facultad de CC. F\'isicas, \\ Universidad Complutense de Madrid, Avda. Complutense s/n, E-28040 Madrid, Spain \\
$^{16}$ Department of Physics, Institute for Astronomy, ETH Z\"urich, CH-8093 Z\"urich, Switzerland \\
$^{17}$ Observatorio Astron\'omico, Laprida 854, X5000BGR, C\'ordoba, Argentina \\
$^{18}$ Consejo de Investigaciones Cient\'ificas y T\'ecnicas de la Rep\'ublica Argentina, Avda. Rivadavia 1917, C1033AAJ, CABA, Argentina \\
$^{19}$ Instituto de Astrof\'isica e Ci\^encias do Espa\c{c}o, Universidade do Porto, CAUP, Rua das Estrelas, P-4150-762 Porto, Portugal \\
$^{20}$ Max Planck Institute for Astronomy, K\"onigstuhl 17, D-69117 Heidelberg, Germany \\
$^{21}$ Leibniz-Institude f\"ur Astrophysik Potsdam (AIP), An der Sternwarte 16, D-14482 Potsdam, Germany \\
$^{22}$ INAF - Osservatorio Astrofisico di Arcetri, Largo Enrico Fermi 5, I-50125 Firenze, Italy
}

\begin{document}

\date{Accepted 2015 October 29.  Received 2015 October 26; in original form 2015 October 1}

\pagerange{\pageref{firstpage}--\pageref{lastpage}} \pubyear{2015}

\maketitle

\label{firstpage}

\begin{abstract}

We study the stellar properties of 44 face-on spiral galaxies from the Calar Alto Legacy Integral Field Area survey via full spectrum fitting techniques. We compare the age profiles with the surface brightness distribution in order to highlight differences between profile types (type I, exponential profile; and II, down-bending profile). We observe an upturn (``U-shape'') in the age profiles for 17 out of these 44 galaxies with reliable stellar information up to their outer parts. This ``U-shape'' is not a unique feature for type II galaxies but can be observed in type I as well. These findings suggest that the mechanisms shaping the surface brightness and stellar population distributions are not directly coupled. This upturn in age is only observable in the light-weighted profiles while it flattens out in the mass-weighted profiles. Given recent results on the outer parts of nearby systems and the results presented in this Letter, one of the most plausible explanations for the age upturn is an early formation of the entire disc ($\sim$~10~Gyr ago) followed by an inside-out quenching of the star formation.

\end{abstract}


\section{Introduction}

The long dynamical time-scales displayed by the stars populating the outer and low surface brightness (SB) regions of disc galaxies make the study of these regions a key element in the understanding of galaxy formation and evolution. Pioneering studies on the light distribution of spiral galaxies found an exponential decline with radius \citep[e.g.][]{1970ApJ...160..811F}. However, deeper and higher-quality images allow us to reach further out. Nowadays, we know that disc galaxies present a wide variety of shapes in the SB profiles of their outer parts \citep[][]{2005ApJ...626L..81E, 2006A&A...454..759P}  showing a continuation of the inner exponential behaviour \citep[type I; e.g.][]{2005ApJ...629..239B}, a lack of light (type II) or an excess of light (type III). Some of the proposed scenarios for the origin of these profiles include a maximum in the angular momentum of the baryonic matter \citep[][]{1987A&A...173...59V}, a star formation (SF) threshold \citep[][]{1989ApJ...344..685K, 2004ApJ...609..667S}, or a combination of both \citep[][]{2001MNRAS.327.1334V}. However, a conclusive explanation is still missing.

Recent theoretical works claim that a substantial number of stars move from their birth location \citep[e.g.][]{2002A&A...388..213B, 2002MNRAS.336..785S, 2012A&A...548A.126M}. This radial migration has implications on the outer parts of type II galaxies. \citet[][]{2008ApJ...675L..65R}, through isolated disc simulations, found that outward migrated stars are the main population in the external parts of disc galaxies as its outer gas surface density is well below the threshold for SF. This change in the stellar populations is responsible for a break in the mass distribution and produces an upturn in the age profile (``U-shape'' age profiles). Cosmological simulations are needed to assess the effect of the environment and satellite accretion in shaping these outer discs \citep[][]{2007ApJ...670..269Y, 2012MNRAS.420..913B}. \citet[][]{2009MNRAS.398..591S}, analysing cosmological simulations, did not find any break in the mass distribution. Even more, the ``U-shape'' age profile was recovered even when radial migration was not allowed. They concluded that the outer ageing of the stellar populations must be due to a radial change in the SF rate caused by a drop in the gas density. Theoretical models for type I and III galaxies are less elaborated and thus, no clear predictions have been stated so far. The analysis of the stellar content in these low density regions in real galaxies is key to refine and constrain galaxy formation models.

Some observational works have studied the stellar content in the outer parts of spiral galaxies. \citet[][]{2008ApJ...683L.103B} stacked 85 $g-r$ colour profiles from \citet[][]{2006A&A...454..759P} separating between type I, II, and III galaxies. They found a clear reddening in the outer parts of their type II galaxies while type I and type III galaxies showed a flattening or a slight blue upturn. This reddening has been also found regardless their SB in \citet[][]{2012ApJ...758...41R}. To minimize the age-metallicity degeneracy that affects colour-based analysis, spectroscopic studies are needed. However, obtaining high quality spectra to analyse the stellar content in these regions is not straightforward. \citet[][]{2012ApJ...752...97Y}, using integral field spectroscopy (IFS) data, examined the radial stellar content of 12 spiral galaxies integrating over elliptical apertures. They were able to reach the outer parts for six type II discs but just three of them displayed the predicted age ``U-shape'' (light-weighted values). This ``U-shape'' implies the presence of old stellar populations in these outer parts regardless of the physical interpretation for such shape.

There is evidence of the existence of this old outer disc  from the analysis of individual stars in very nearby systems \citep[e.g.][]{2012MNRAS.420.2625B, 2015MNRAS.446.2789B, 2012ApJ...753..138R}. In addition, recent studies have observed such behaviour for galactic systems regardless of their morphology, formation, or mass. \citet[][]{2012AJ....143...47Z} analysed 34 dwarf irregular galaxies finding hints of old stellar populations over the entire disc from their spectral energy distribution (SED) modelling. Similar results have been found by \citet[][]{2015ApJ...800..120Z} analysing SED modelling of 698 galaxies from the Pan-STARRS1 Medium Deep Survey images. 

To date, there are no clear observational evidences of a correlation between SB and other stellar parameters such as age. Here, we present for the first time the radial behaviour of the age of the stellar populations (light- and mass-weighted) up to the outer parts of a large sample of spiral galaxies from the Calar Alto Legacy Integral Field Area survey \citep[CALIFA;][]{2012A&A...538A...8S}, focusing on the relation between the age distribution in the outer parts and their SB profiles.

\section[data]{Sample and CALIFA data}
\label{data}

In this work, we make use of the IFS data from the CALIFA survey \citep[][]{2012A&A...538A...8S}. CALIFA data were obtained at the 3.5\,m telescope at Calar Alto using the PMAS/PPAK spectrograph. This project provides high quality spectra over a wide field-of-view (FoV) of 72 arcsec$\times$64 arcsec of $\sim$ 600 galaxies in the local Universe \citep{2014A&A...569A...1W} with two different observing set-ups: one at higher resolution (V1200); and the other at a lower resolution (V500). The wavelength range of the V500 (V1200) data is 3745 -- 7500 \AA~(3650 -- 4840 \AA) with a spectral resolution of full width at half-maximum (FWHM) = 6.0 \AA $ $ (FWHM = 2.7 \AA). For more information about the CALIFA data see \citet{2015A&A...576A.135G}. The wide FoV of the instrument and the thoroughly tested sky subtraction procedure implemented in the reduction pipeline allow us to analyse the stellar and gas content up to the outer parts \citep[][]{2013A&A...549A..87H, 2015arXiv150907878M}. We consider ``outer parts'' of our galaxies those beyond the break radius for type II and beyond three disc-scalelengths for type I. We use a combination of the V500 and the V1200 (degraded to the V500 resolution) data cubes (v1.5) in order to expand the wavelength range to 3700 -- 7500 \AA~(COMBO cube).

We choose a subsample of 44 low-to-intermediate-inclined ($0^\circ\leq i\leq 75^\circ$), small ($d_{25}<$94.8 arcsec), non-interacting, spiral ($0\leq T\leq 8$) galaxies observed by CALIFA based on the Hyperleda catalogue. Observational constraints such as high S/N in the outer parts or enough instrument FoV have been taken into account to ensure a reliable recovery of the stellar content up to the outer parts.

We have carefully analysed the two dimensional light distribution in our galaxies to characterize their morphology and SB profiles. We use the fully-calibrated Sloan Digital Sky Survey (SDSS) images ($g$, $r$, and $i$ bands) from the seventh data release \citep[DR7;][]{2009ApJS..182..543A} and the GASP2D software \citep[GAlaxy Surface Photometry 2 Dimensional Decomposition;][]{2008A&A...478..353M}. GASP2D adopts a pixel-by-pixel Levenberg-Marquardt algorithm to fit the two-dimensional SB distribution of the galaxy allowing us to model different structural galaxy components such as bulge, disc (broken disc), or bar. All this two-dimensional analysis will be presented in detail in M\'endez-Abreu et al. (in preparation).

\begin{table*}
{\normalsize
\centering
\begin{tabular}{llllllllll}
\hline\hline
Galaxy & RA & Dec. & Morphological & Bar & SB & $h_{\rm in}$ & $h_{\rm out}$ & $R_{\rm break}/h_{\rm in}$ & $R_{\rm min}$ \\ 
 & (h:m:s) & ($^o$ $'$ $''$) & type & (1, yes; 0 no) & profile & $(kpc)$ & $(kpc)$ & $ $ & $(kpc)$  \\ 
$(1)$ & $(2)$ & $(3)$ & $(4)$ & $(5)$ & $(6)$ & $(7)$ & $(8)$ & $(9)$ & $(10)$\\ \hline
 IC1199      & 16:10:34.34  &  +10:02:25.32  & Sbc & 0 &  II &   5.83 &  2.54 & 1.46 & 7.00  \\
 NGC0234     & 00:43:32.39  &  +14:20:33.24  & SABc& 0 &  II &   5.56 &  3.11 & 1.55 & 6.67  \\
 NGC0551     & 01:27:40.63  &  +37:10:58.73  & SBbc& 1 &   I &   5.09 &   -   & -  & 10.18  \\  
 NGC2449     & 07:47:20.29  &  +26:55:48.70  & Sab & 1 &   I &   4.50 &   -   & -  & 6.53  \\  
 NGC2572     & 08:21:24.62  &  +19:08:51.99  & Sa  & 1 &   I &   4.79 &   -   & -  & 8.62  \\  
 NGC3815     & 11:41:39.29  & +24:48:01.79   & Sab & 1 &   I &   2.83 &   -   & -  & 7.36  \\
 NGC4470     & 12:29:37.77  & +07:49:27.12   & Sa  & 0 &   I &   1.76 &   -   & -  & 2.20  \\
 NGC4711     & 12:48:45.86  &  +35:19:57.72  & SBb & 0 &  II &   3.94 &  2.39 & 1.85 & 6.42  \\
 NGC5376     & 13:55:16.05  &  +59:30:23.80  & SABa& 0 &  II &   2.95 &  1.69 & 1.50 & 3.54  \\  
 NGC5633     & 14:27:28.38  &  +46:08:47.67  & Sb  & 0 &   I &   1.43 &   -   & -  & 2.43  \\  
 NGC5732     & 14:40:38.95  &  +38:38:16.16  & Sbc & 0 &  II &   2.58 &  1.50 & 3.75 & 3.61  \\
 NGC6155     & 16:26:08.33  &  +48:22:00.46  & Sc  & 0 &  II &   1.95 &  1.22 & 3.27 & 3.02  \\
 NGC6478     & 17:48:37.73  &  +51:09:13.68  & Sc  & 0 &  II &   9.57 &  5.50 & 1.31 & 15.31  \\
 NGC7311     & 22:34:06.79  &  +05:34:13.16  & Sab & 0 &   I &   3.63 &   -   & -  & 7.44  \\   
 NGC7321     & 22:36:28.02  &  +21:37:18.35  & Sbc & 1 &  II &  44.69 & 4.95 & 0.32 & 12.07  \\
 UGC00036    & 00:05:13.88  &  +06:46:19.30  & Sa  & 1 &   I &   4.08 &   -   & -  & 8.57  \\  
 UGC05396    & 10:01:40.48  &  +10:45:23.13  & Sc  & 1 &  II &   6.38 &  2.65 & 2.67 & 10.85  \\ \hline
\end{tabular}
\caption{List of galaxies displaying ``U-shape'' age profiles. (1) Name of the galaxy; (2) right ascension (J2000); (3) declination (J2000); (4) morphological type; (5) bar (1, yes; 0, no); (6) surface brightness profile type; (7) inner disc scalelength ($h_{\rm in}$ in kpc); (8) outer disc scalelength ($h_{\rm out}$ in kpc); (9) break radius ($R_{\rm break}$) in units of $h_{\rm in}$; (10) minimum age position ($R_{\rm min}$ in kpc). Columns (1)--(4) from the hyperleda data base. Columns (5)--(9) from our 2D decomposition.} 
\label{galaxy_tab}
}
\end{table*}

\section[analysis]{Analysis}
\label{analysis}

In this section we describe the procedure followed to analyse the stellar content from the CALIFA data. A complete description of the analysis and the sample of galaxies will be given in Ruiz-Lara et al. (in preparation).

\begin{figure}
\centering 
\includegraphics[scale=0.39]{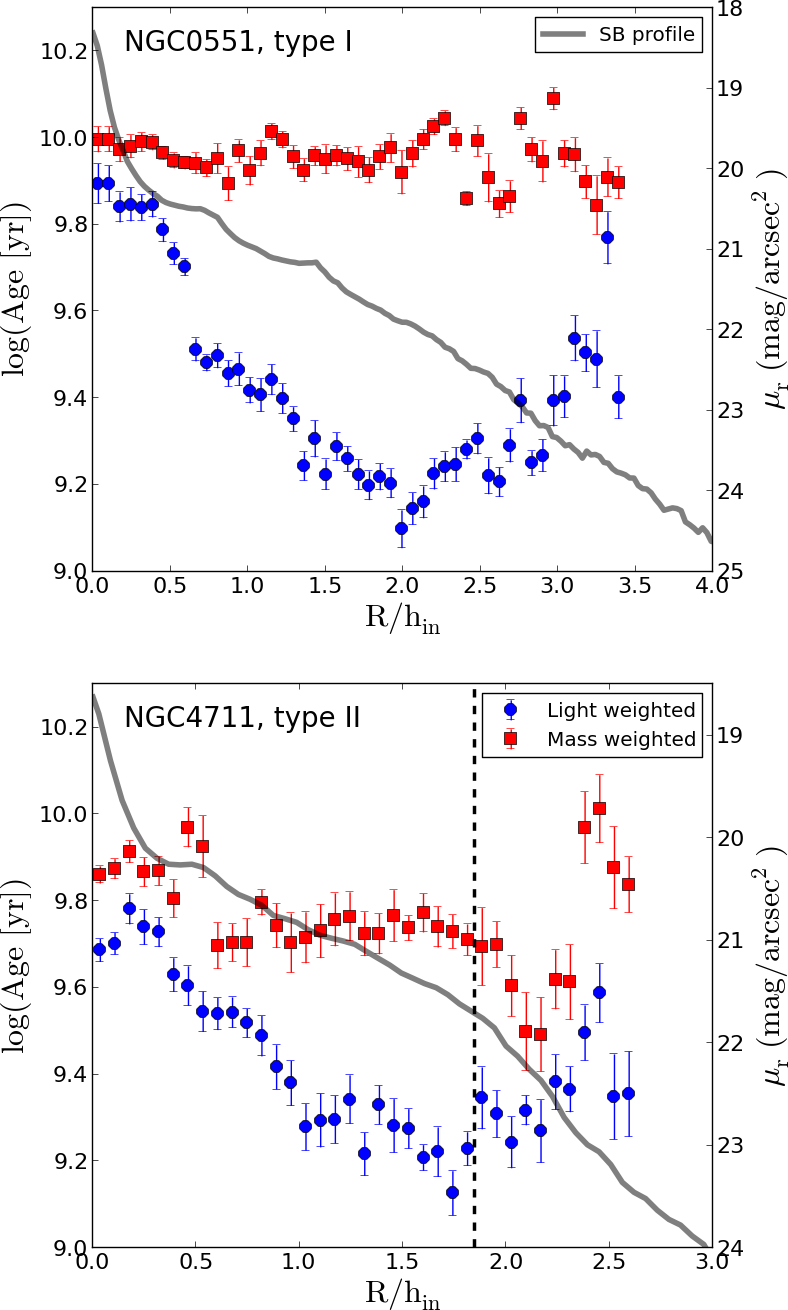} \\   
\caption{Age radial profile in logarithmic scale and surface brightness profile (SDSS $r$-band) for NGC~0551 (top panel, type I) and NGC~4711 (bottom panel, type II). Red (blue) squares (points) indicate mass-weighted (light-weighted) values. The black dashed line is located at the break radius. Grey continuous lines corresponds to the SB profiles (see right {\it y}-axis).} 
\label{ages_plot} 
\end{figure}

(i) We apply a stellar kinematic extraction method designed for dealing with CALIFA data (Falc\'on-Barroso et al. in preparation) to correct the observed datacubes for the stellar kinematics effect. After applying an adaptive Voronoi method following \citet[][]{2003MNRAS.342..345C} with a goal S/N of 20 and considering only spaxels with continuum S/N $>$ 3, we use the penalized pixel fitting code \citep[{\tt pPXF};][]{2004PASP..116..138C, 2011MNRAS.413..813C} to extract the stellar kinematics.

(ii) We perform a radial elliptical binning to the kinematic-corrected data (rest frame and a common velocity dispersion). We use different steps from one ellipse to the next in order to reach a S/N $>$ 20 in the spectra (per \AA). The centre, ellipticity, and position angle of the ellipses are fixed, matching the outer disc isophotes.

(iii) We use {\tt GANDALF} \citep[Gas AND Absorption Line Fitting;][]{2006MNRAS.366.1151S, 2006MNRAS.369..529F} to subtract the emission line component from the observed spectra. We adopt an optimal subset of the \citet{2010MNRAS.404.1639V} models (hereafter MILES models) as stellar templates. We use 30 log-spaced age bins (from 0.063 to 17.8 Gyr) and all {\tt MILES} metallicities. 

(iv) The stellar population parameters are derived using STEllar Content and Kinematics via Maximum A Posteriori likelihood \citep[{\tt STECKMAP};][]{2006MNRAS.365...74O, 2006MNRAS.365...46O}. {\tt STECKMAP} is aimed at simultaneously recovering the stellar content and kinematics using a Bayesian method via a maximum a posteriori algorithm. It is based on the minimization of a penalized $\chi^2$ while no a priori shape of the solution is assumed (i.e. it is a non-parametric code). See \citet[][]{2011MNRAS.415..709S, 2014A&A...570A...6S} for further details. To compute the age, luminosity or mass weighted (Q-W), we use the following equation:
\begin{equation}
  <{\rm log}({\rm Age} [{\rm yr}])>_{{\rm Q-W}} = {\sum_i Q(i)*{\rm log}({\rm Age}_i) \over \sum_i Q(i)},
\end{equation}
where $Q(i)$ is the light or mass fraction of each age bin ($i$). We fix the stellar kinematics in order to avoid the velocity dispersion-metallicity degeneracy \citep[][]{2011MNRAS.415..709S}. Statistical errors are computed by means of 25 Monte Carlo simulations as computed by {\tt STECKMAP}. Once {\tt STECKMAP} has determined the best combination of model templates \citep[][]{2010MNRAS.404.1639V} to fit the observed spectrum, we add noise based on the spectrum S/N and run {\tt STECKMAP} again. This test is done 25 times and the standard deviation of the recovered age is considered as the error. Along with the age profiles, metallicity profiles are obtained. We must warn the reader that, although we have extensively tested our method  \citep[][]{2015A&A...583A..60R} against the well known age--metallicity degeneracy using Monte Carlo simulations finding no clear trends, some effects might still remain.

\section[results]{Results \& discussion}
\label{results}

\begin{figure} 
\centering 
\includegraphics[scale=0.44]{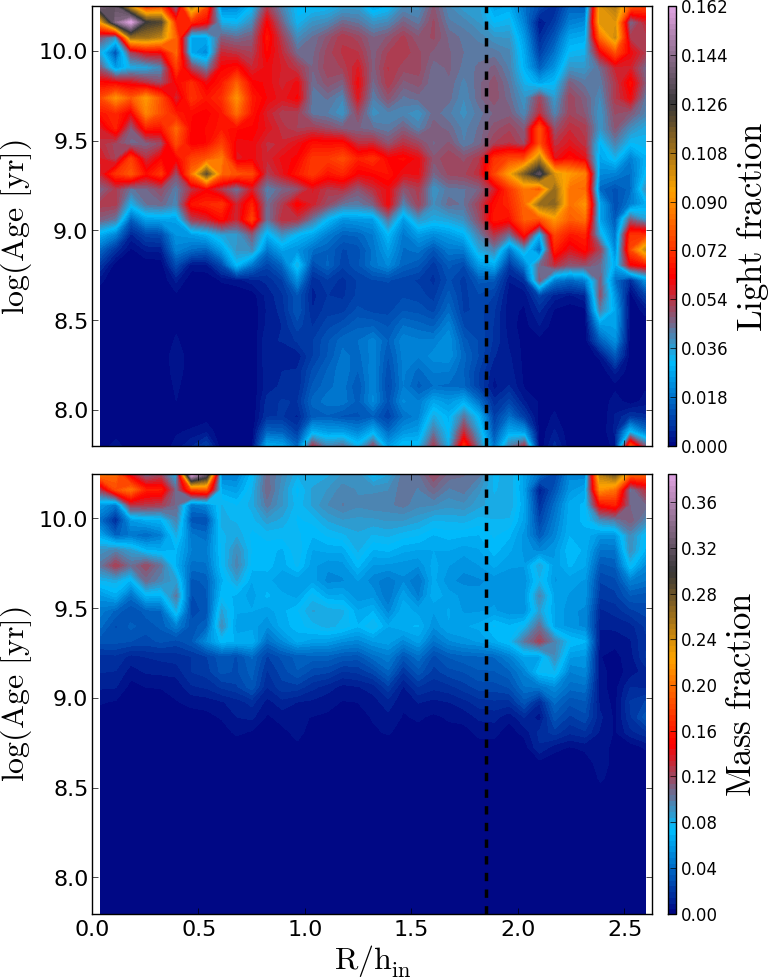} \\   
\caption{Radial stellar age distribution for NGC~4711: Upper panel, luminosity-weighted; bottom panel, mass-weighted. The black dashed line is located at the break radius.} 
\label{sfh_plot} 
\end{figure}

We study the radial distribution of stellar parameters up to the outer parts (i.e. galaxies for which we reach beyond their break radius or at least $\sim$ 3 disc-scalelengths) and relate these properties to their SB profiles for a sample of 44 galaxies (see Sect.~\ref{data}). Our selection criteria leave us with no galaxies displaying a type III SB profile, so our results are statistically limited to type I and II galaxies.

We find that 17 galaxies (8 type I and 9 type II galaxies), i.e. 39\% (40\% of the type I and 38\% of the type II galaxies) of our final sample show clear ``U-shapes'' in the luminosity-weighted (L-W) age profile regardless their SB distribution (Table~\ref{galaxy_tab}). Such ``U-shape'' disappears for all cases when the age is mass-weighted (M-W), displaying a flat distribution in the whole disc ($\sim$~10~Gyr) with a slight correlation with morphological type in the sense of what found in \citet[][]{2015A&A...581A.103G} but to a lesser degree. The remaining 27 galaxies present different behaviours with no clear pattern. Figure~\ref{ages_plot} shows the age profiles for two galaxies as examples of a type I (NGC~0551) and a type II (NGC~4711).

For the first time, ``U-shaped'' age profiles are found indistinctly in type I and II galaxies from spectroscopic data. To date, such profiles have been found (observationally and theoretically) just in type II galaxies \citep[][]{2008ApJ...675L..65R, 2008ApJ...683L.103B, 2009MNRAS.398..591S, 2012ApJ...752...97Y}. However, recent cosmological simulations (Ruiz-Lara et al., submitted) analysing 19 Milky Way-like galaxies with different SB profiles from the {\tt RaDES} set of galaxies \citep[][]{2012A&A...547A..63F} show this upturn in age for all of them (in agreement with \citealt[][]{2015ApJ...804L...9M}). An important result from this work in constraining galaxy formation models is the fact that not all the analysed galaxies display a ``U-shape'' in the age profile and it appears in both type I and II. These findings allow us to conclude that the mechanisms responsible for shaping the SB distribution are not coupled to those shaping the stellar age profile. Thus, we can rule out that breaks are linked to particular changes in the stellar populations. In other words, old outer discs do not lead to type II SB profiles as proposed by numerical simulations. Other aspects must be responsible for the observed SB profiles \citep[e.g.][]{2015MNRAS.448L..99H}.

It is important to note that the ``U-shape'' age profiles shown in simulations are M-W quantities. However, our M-W profiles show no sign of ``U-shape'' at all, meaning that SF in simulations has been higher at later times than in real galaxies. We also tested if there is any correlation between the ``U-shape'' occurrence with morphology, galaxy mass, and bar presence. We do not find any pattern among these properties, although we suffer from low number statistics and these relations should be further investigated. 

In most cases, the observed minimum age is located prior to the break radius, with the exceptions of NGC~6478 and NGC~4711. There is no consensus about a possible link between the location of the break in the light distribution and the minimum in age, neither from observations nor from theoretical works. While some works point towards younger stars (bluer colours) located exactly at the break radius \citep[][]{2008ApJ...675L..65R, 2008ApJ...683L.103B, 2015arXiv150907878M} others claim that such minimum age is found prior to it \citep[][]{2009MNRAS.398..591S, 2012ApJ...752...97Y} or with no relation at all \citep[][]{2012ApJ...758...41R}. From this work we can rule out that the minimum in age is exactly located at the break radius.

L-W radial stellar age distributions (see example in left-hand panel of Fig.~\ref{sfh_plot}) for the 44 galaxies show clear evidences suggesting that SF in them have been quenched inside-out (i.e. inner regions mainly dominated by old stars, with young stars becoming more important as we move outwards), in agreement with previous CALIFA works on the stellar mass growth \citep[][]{2013ApJ...764L...1P}, stellar populations radial profiles \citep[][]{2014A&A...562A..47G, 2014A&A...570A...6S, 2015A&A...581A.103G} and gas abundance gradients \citep[][S\'anchez-Menguiano et al., submitted]{2012A&A...546A...2S, 2014A&A...563A..49S}. Apart from that, an outermost old stellar component appears again for those galaxies presenting ``U-shape''. The M-W radial stellar age distributions (see example in right-hand panel of Fig.~\ref{sfh_plot}) do not show any population younger than 2~Gyr with the bulk of stars being $\sim$~10~Gyr old. 

This work supports recent observations finding old outer discs \citep[e.g.][]{2012MNRAS.420.2625B, 2015MNRAS.446.2789B, 2012ApJ...753..138R}, pointing towards an early SF along the entire disc followed by an inside-out SF quenching as the cause of the age upturn. However, we cannot rule out scenarios considering the effects of other mechanisms (radial migration, different SF recipes, satellite accretion, etc.) shaping different SB profiles after the age profile is built.

\section{Conclusions}

In this Letter, we study the stellar age profiles up to the outer parts of the discs of a large sample of galaxies selected from CALIFA. We conclude that (i) mechanisms shaping the SB and stellar distributions are not coupled, i.e. more processes must be involved in the type II SB profile formation; (ii) ``U-shape'' age profiles are not unique for type II discs; (iii) this upturn is not a universal feature (appears only in 17 out of 44 galaxies). A possible explanation for the age upturn compatible with our results is an early build-up of the entire disc followed by an inside-out SF quenching.

\section*{Acknowledgements}

This Letter is based on data from the CALIFA Survey (\url{http://califa.caha.es}), funded by the Spanish Ministry of Science (grant ICTS-2009-10), and the Centro Astron\'omico Hispano-Alem\'an. This research has been partly supported by the Spanish Ministry of Science and Innovation under grants AYA2011-24728, AYA2013-48226-C3-1-P and Consolider-Ingenio CSD2010-00064; and by the Junta de Andaluc\'ia (FQM-108). TRL thanks the support of the Spanish Ministerio de Educaci\'on, Cultura y Deporte (FPU fellowship). We acknowledge the usage of the HyperLeda data base (\url{http://leda.univ-lyon1.fr}) and SDSS data (\url{http://www.sdss.org/collaboration/citing-sdss/}).

\bibliography{bibliography}

\begin{thebibliography}{49}
\expandafter\ifx\csname natexlab\endcsname\relax\def\natexlab#1{#1}\fi

\bibitem[{{Abazajian} {et~al}\mbox{.}(2009){Abazajian}
  {et~al.}}]{2009ApJS..182..543A}
{Abazajian} K.~N., {et~al.}, 2009, \apjs, 182, 543

\bibitem[{{Bakos}, {Trujillo} \& {Pohlen}(2008){Bakos}, {Trujillo}, \&
  {Pohlen}}]{2008ApJ...683L.103B}
{Bakos} J., {Trujillo} I., {Pohlen} M., 2008, \apjl, 683, L103

\bibitem[{{Battaner}, {Florido} \& {Jim{\'e}nez-Vicente}(2002){Battaner},
  {Florido}, \& {Jim{\'e}nez-Vicente}}]{2002A&A...388..213B}
{Battaner} E., {Florido} E., {Jim{\'e}nez-Vicente} J., 2002, \aap, 388, 213

\bibitem[{{Bernard} {et~al}\mbox{.}(2012){Bernard}
  {et~al.}}]{2012MNRAS.420.2625B}
{Bernard} E.~J., {et~al.}, 2012, \mnras, 420, 2625

\bibitem[{{Bernard} {et~al}\mbox{.}(2015){Bernard}
  {et~al.}}]{2015MNRAS.446.2789B}
{Bernard} E.~J., {et~al.}, 2015, \mnras, 446, 2789

\bibitem[{{Bird}, {Kazantzidis} \& {Weinberg}(2012){Bird}, {Kazantzidis}, \&
  {Weinberg}}]{2012MNRAS.420..913B}
{Bird} J.~C., {Kazantzidis} S., {Weinberg} D.~H., 2012, \mnras, 420, 913

\bibitem[{{Bland-Hawthorn} {et~al}\mbox{.}(2005){Bland-Hawthorn}, {Vlaji{\'c}},
  {Freeman}, \& {Draine}}]{2005ApJ...629..239B}
{Bland-Hawthorn} J., {Vlaji{\'c}} M., {Freeman} K.~C., {Draine} B.~T., 2005,
  \apj, 629, 239

\bibitem[{{Cappellari} \& {Copin}(2003)}]{2003MNRAS.342..345C}
{Cappellari} M., {Copin} Y., 2003, \mnras, 342, 345

\bibitem[{{Cappellari} \& {Emsellem}(2004)}]{2004PASP..116..138C}
{Cappellari} M., {Emsellem} E., 2004, \pasp, 116, 138

\bibitem[{{Cappellari} {et~al}\mbox{.}(2011){Cappellari}
  {et~al.}}]{2011MNRAS.413..813C}
{Cappellari} M., {et~al.}, 2011, \mnras, 413, 813

\bibitem[{{Erwin}, {Beckman} \& {Pohlen}(2005){Erwin}, {Beckman}, \&
  {Pohlen}}]{2005ApJ...626L..81E}
{Erwin} P., {Beckman} J.~E., {Pohlen} M., 2005, \apjl, 626, L81

\bibitem[{{Falc{\'o}n-Barroso} {et~al}\mbox{.}(2006){Falc{\'o}n-Barroso}
  {et~al.}}]{2006MNRAS.369..529F}
{Falc{\'o}n-Barroso} J., {et~al.}, 2006, \mnras, 369, 529

\bibitem[{{Few} {et~al}\mbox{.}(2012){Few}, {Gibson}, {Courty},
  {Michel-Dansac}, {Brook}, \& {Stinson}}]{2012A&A...547A..63F}
{Few} C.~G., {Gibson} B.~K., {Courty} S., {Michel-Dansac} L., {Brook} C.~B.,
  {Stinson} G.~S., 2012, \aap, 547, A63

\bibitem[{{Freeman}(1970)}]{1970ApJ...160..811F}
{Freeman} K.~C., 1970, \apj, 160, 811

\bibitem[{{Garc{\'{\i}}a-Benito} {et~al}\mbox{.}(2015){Garc{\'{\i}}a-Benito}
  {et~al.}}]{2015A&A...576A.135G}
{Garc{\'{\i}}a-Benito} R., {et~al.}, 2015, \aap, 576, A135

\bibitem[{{Gonz{\'a}lez Delgado} {et~al}\mbox{.}(2014){Gonz{\'a}lez Delgado}
  {et~al.}}]{2014A&A...562A..47G}
{Gonz{\'a}lez Delgado} R.~M., {et~al.}, 2014, \aap, 562, A47

\bibitem[{{Gonz{\'a}lez Delgado} {et~al}\mbox{.}(2015){Gonz{\'a}lez Delgado}
  {et~al.}}]{2015A&A...581A.103G}
{Gonz{\'a}lez Delgado} R.~M., {et~al.}, 2015, \aap, 581, A103

\bibitem[{{Herpich} {et~al}\mbox{.}(2015){Herpich}
  {et~al.}}]{2015MNRAS.448L..99H}
{Herpich} J., {et~al.}, 2015, \mnras, 448, L99

\bibitem[{{Husemann} {et~al}\mbox{.}(2013){Husemann}
  {et~al.}}]{2013A&A...549A..87H}
{Husemann} B., {et~al.}, 2013, \aap, 549, A87

\bibitem[{{Kennicutt}(1989)}]{1989ApJ...344..685K}
{Kennicutt}, Jr. R.~C., 1989, \apj, 344, 685

\bibitem[{{Marino} {et~al}\mbox{.}(2015){Marino}
  {et~al.}}]{2015arXiv150907878M}
{Marino} R.~A., {et~al.}, 2015, ArXiv e-prints

\bibitem[{{M{\'e}ndez-Abreu} {et~al}\mbox{.}(2008){M{\'e}ndez-Abreu},
  {Aguerri}, {Corsini}, \& {Simonneau}}]{2008A&A...478..353M}
{M{\'e}ndez-Abreu} J., {Aguerri} J.~A.~L., {Corsini} E.~M., {Simonneau} E.,
  2008, \aap, 478, 353

\bibitem[{{Minchev} {et~al}\mbox{.}(2012){Minchev}, {Famaey}, {Quillen}, {Di
  Matteo}, {Combes}, {Vlaji{\'c}}, {Erwin}, \&
  {Bland-Hawthorn}}]{2012A&A...548A.126M}
{Minchev} I., {Famaey} B., {Quillen} A.~C., {Di Matteo} P., {Combes} F.,
  {Vlaji{\'c}} M., {Erwin} P., {Bland-Hawthorn} J., 2012, \aap, 548, A126

\bibitem[{{Minchev} {et~al}\mbox{.}(2015){Minchev}, {Martig}, {Streich},
  {Scannapieco}, {de Jong}, \& {Steinmetz}}]{2015ApJ...804L...9M}
{Minchev} I., {Martig} M., {Streich} D., {Scannapieco} C., {de Jong} R.~S.,
  {Steinmetz} M., 2015, \apjl, 804, L9

\bibitem[{{Ocvirk} {et~al}\mbox{.}(2006{\natexlab{a}}){Ocvirk}, {Pichon},
  {Lan{\c c}on}, \& {Thi{\'e}baut}}]{2006MNRAS.365...74O}
{Ocvirk} P., {Pichon} C., {Lan{\c c}on} A., {Thi{\'e}baut} E.,
  2006{\natexlab{a}}, \mnras, 365, 74

\bibitem[{{Ocvirk} {et~al}\mbox{.}(2006{\natexlab{b}}){Ocvirk}, {Pichon},
  {Lan{\c c}on}, \& {Thi{\'e}baut}}]{2006MNRAS.365...46O}
{Ocvirk} P., {Pichon} C., {Lan{\c c}on} A., {Thi{\'e}baut} E.,
  2006{\natexlab{b}}, \mnras, 365, 46

\bibitem[{{P{\'e}rez} {et~al}\mbox{.}(2013){P{\'e}rez}
  {et~al.}}]{2013ApJ...764L...1P}
{P{\'e}rez} E., {et~al.}, 2013, \apjl, 764, L1

\bibitem[{{Pohlen} \& {Trujillo}(2006)}]{2006A&A...454..759P}
{Pohlen} M., {Trujillo} I., 2006, \aap, 454, 759

\bibitem[{{Radburn-Smith} {et~al}\mbox{.}(2012){Radburn-Smith}
  {et~al.}}]{2012ApJ...753..138R}
{Radburn-Smith} D.~J., {et~al.}, 2012, \apj, 753, 138

\bibitem[{{Roediger} {et~al}\mbox{.}(2012){Roediger}, {Courteau},
  {S{\'a}nchez-Bl{\'a}zquez}, \& {McDonald}}]{2012ApJ...758...41R}
{Roediger} J.~C., {Courteau} S., {S{\'a}nchez-Bl{\'a}zquez} P., {McDonald} M.,
  2012, \apj, 758, 41

\bibitem[{{Ro{\v s}kar} {et~al}\mbox{.}(2008){Ro{\v s}kar}, {Debattista},
  {Stinson}, {Quinn}, {Kaufmann}, \& {Wadsley}}]{2008ApJ...675L..65R}
{Ro{\v s}kar} R., {Debattista} V.~P., {Stinson} G.~S., {Quinn} T.~R.,
  {Kaufmann} T., {Wadsley} J., 2008, \apjl, 675, L65

\bibitem[{{Ruiz-Lara} {et~al}\mbox{.}(2015){Ruiz-Lara}
  {et~al.}}]{2015A&A...583A..60R}
{Ruiz-Lara} T., {et~al.}, 2015, \aap, 583, A60

\bibitem[{{S{\'a}nchez} {et~al}\mbox{.}(2012{\natexlab{a}}){S{\'a}nchez}
  {et~al.}}]{2012A&A...538A...8S}
{S{\'a}nchez} S.~F., {et~al.}, 2012{\natexlab{a}}, \aap, 538, A8

\bibitem[{{S{\'a}nchez} {et~al}\mbox{.}(2012{\natexlab{b}}){S{\'a}nchez}
  {et~al.}}]{2012A&A...546A...2S}
{S{\'a}nchez} S.~F., {et~al.}, 2012{\natexlab{b}}, \aap, 546, A2

\bibitem[{{S{\'a}nchez} {et~al}\mbox{.}(2014){S{\'a}nchez}
  {et~al.}}]{2014A&A...563A..49S}
{S{\'a}nchez} S.~F., {et~al.}, 2014, \aap, 563, A49

\bibitem[{{S{\'a}nchez-Bl{\'a}zquez}
  {et~al}\mbox{.}(2009){S{\'a}nchez-Bl{\'a}zquez}, {Courty}, {Gibson}, \&
  {Brook}}]{2009MNRAS.398..591S}
{S{\'a}nchez-Bl{\'a}zquez} P., {Courty} S., {Gibson} B.~K., {Brook} C.~B.,
  2009, \mnras, 398, 591

\bibitem[{{S{\'a}nchez-Bl{\'a}zquez}
  {et~al}\mbox{.}(2011){S{\'a}nchez-Bl{\'a}zquez}, {Ocvirk}, {Gibson},
  {P{\'e}rez}, \& {Peletier}}]{2011MNRAS.415..709S}
{S{\'a}nchez-Bl{\'a}zquez} P., {Ocvirk} P., {Gibson} B.~K., {P{\'e}rez} I.,
  {Peletier} R.~F., 2011, \mnras, 415, 709

\bibitem[{{S{\'a}nchez-Bl{\'a}zquez}
  {et~al}\mbox{.}(2014){S{\'a}nchez-Bl{\'a}zquez}
  {et~al.}}]{2014A&A...570A...6S}
{S{\'a}nchez-Bl{\'a}zquez} P., {et~al.}, 2014, \aap, 570, A6

\bibitem[{{Sarzi} {et~al}\mbox{.}(2006){Sarzi} {et~al.}}]{2006MNRAS.366.1151S}
{Sarzi} M., {et~al.}, 2006, \mnras, 366, 1151

\bibitem[{{Schaye}(2004)}]{2004ApJ...609..667S}
{Schaye} J., 2004, \apj, 609, 667

\bibitem[{{Sellwood} \& {Binney}(2002)}]{2002MNRAS.336..785S}
{Sellwood} J.~A., {Binney} J.~J., 2002, \mnras, 336, 785

\bibitem[{{van den Bosch}(2001)}]{2001MNRAS.327.1334V}
{van den Bosch} F.~C., 2001, \mnras, 327, 1334

\bibitem[{{van der Kruit}(1987)}]{1987A&A...173...59V}
{van der Kruit} P.~C., 1987, \aap, 173, 59

\bibitem[{{Vazdekis} {et~al}\mbox{.}(2010){Vazdekis},
  {S{\'a}nchez-Bl{\'a}zquez}, {Falc{\'o}n-Barroso}, {Cenarro}, {Beasley},
  {Cardiel}, {Gorgas}, \& {Peletier}}]{2010MNRAS.404.1639V}
{Vazdekis} A., {S{\'a}nchez-Bl{\'a}zquez} P., {Falc{\'o}n-Barroso} J.,
  {Cenarro} A.~J., {Beasley} M.~A., {Cardiel} N., {Gorgas} J., {Peletier}
  R.~F., 2010, \mnras, 404, 1639

\bibitem[{{Walcher} {et~al}\mbox{.}(2014){Walcher}
  {et~al.}}]{2014A&A...569A...1W}
{Walcher} C.~J., {et~al.}, 2014, \aap, 569, A1

\bibitem[{{Yoachim}, {Ro{\v s}kar} \& {Debattista}(2012){Yoachim}, {Ro{\v
  s}kar}, \& {Debattista}}]{2012ApJ...752...97Y}
{Yoachim} P., {Ro{\v s}kar} R., {Debattista} V.~P., 2012, \apj, 752, 97

\bibitem[{{Younger} {et~al}\mbox{.}(2007){Younger}, {Cox}, {Seth}, \&
  {Hernquist}}]{2007ApJ...670..269Y}
{Younger} J.~D., {Cox} T.~J., {Seth} A.~C., {Hernquist} L., 2007, \apj, 670,
  269

\bibitem[{{Zhang} {et~al}\mbox{.}(2012){Zhang}, {Hunter}, {Elmegreen}, {Gao},
  \& {Schruba}}]{2012AJ....143...47Z}
{Zhang} H.-X., {Hunter} D.~A., {Elmegreen} B.~G., {Gao} Y., {Schruba} A., 2012,
  \aj, 143, 47

\bibitem[{{Zheng} {et~al}\mbox{.}(2015){Zheng} {et~al.}}]{2015ApJ...800..120Z}
{Zheng} Z., {et~al.}, 2015, \apj, 800, 120

\end{thebibliography}

\bsp

\end{document}